\documentclass[10 pt, conference, letterpaper]{IEEEtran}
\usepackage{times}
\usepackage{verbatim}
\usepackage{cite}
\usepackage{graphicx}
\usepackage{color}
\usepackage{authblk}
\usepackage{commath}
\usepackage{tabulary}
\graphicspath{ {Experimental_Results/} }

\usepackage[figurename=Figure,tablename=Table]{caption}
\usepackage{xr}
\providecommand{\keywords}[1]{\textbf{\textit{Keywords---}}#1}

\author {
	{{Pankaj Saha},{ Angel Beltre}, and {Madhusudhan Govindaraju}}
	\vspace{1.6mm}\\
	\fontsize{10}{10}\selectfont\itshape
    {Cloud and Big Data Laboratory},
	{State University of New York (SUNY) at Binghamton}\\
	\fontsize{9}{9}\selectfont\ttfamily\upshape
	{
		{{$\lbrace$psaha4, abeltre1, mgovinda$\rbrace$@binghamton.edu }}
	}
}

\begin{document}

\def\sharedaffiliation{%
\end{tabular}
\begin{tabular}{c}}

\title{Scylla: A Mesos Framework for Container Based MPI Jobs}

\renewcommand{\thetable}{\arabic{table}}
\date{}
\maketitle
\thispagestyle{empty}
\pagestyle{empty}

\begin{abstract}
Open source cloud technologies provide a wide range of support for
creating customized compute node clusters to schedule tasks and
managing resources. In cloud infrastructures such as
Jetstream and Chameleon, which are used for scientific research, users
receive complete control of the Virtual Machines (VM) that are allocated to
them. Importantly, users get root access to the VMs. This provides an
opportunity for HPC users to experiment with new resource management
technologies such as Apache Mesos that have proven scalability,
flexibility, and fault tolerance. To ease the development and
deployment of HPC tools on the cloud, the containerization technology
has matured and is gaining interest in the scientific community. In
particular, several well known scientific code bases now have publicly
available Docker containers. While Mesos provides support for Docker
containers to execute individually, it does not provide support for
container inter-communication or orchestration of the containers for a
parallel or distributed application.  In this paper, we present the
design, implementation, and performance analysis of a Mesos framework,
{\it Scylla}, which integrates Mesos with Docker Swarm to enable
orchestration of MPI jobs on a cluster of VMs acquired from the
Chameleon cloud\cite{ChameleonCloud}. Scylla uses Docker Swarm for communication between
containerized tasks (MPI processes) and Apache Mesos for resource
pooling and allocation. Scylla allows a policy driven approach to
determine how the containers should be distributed across the nodes
depending on the CPU, memory, and network throughput requirement for
each application.
\end{abstract}

\section{Introduction}

The Jetstream \cite{Jetstream:Cloud} \cite{Stewart2015Jetstream} and Chameleon \cite{ChameleonCloud} cloud infrastructures
provide a platform for the research community to explore new solutions
for several scientific computing challenges including new software
architectures for cluster computing. These cloud infrastructures
provide users or communities with VMs and bare metal nodes that have
varied resource configurations in terms of numbers of cores, memory capacity, network and I/O bandwidth. Users have complete control
over the VMs, including root access to the VMs. Additionally, users
also have the flexibility to choose their tools for scheduling,
resource management, fault tolerance, and other cluster management
tasks. While the scientific community has so far primarily used tools
such as Slurm\cite{SlurmManager} and Torque\cite{TORQUEManager} on bare metal clusters, there is an
opportunity to leverage the open source Big Data and Cloud tools that
have several desired features including performance, fault tolerance,
and massive scalability.

Open source tools, especially the ones belonging to the Apache family, are
being widely used in production settings. Interestingly, they are
receiving active contributions from experts in several companies in
the Big Data and Cloud space. In particular, Apache Mesos~\cite{Hindman2011Mesos:Center} has
emerged as a leader for large-scale cluster management. As of now, more than 100 companies use Mesos for their software infrastructure. In companies such as Apple and Twitter, it is estimated to be deployed on more than 10K nodes\cite{MesosNews}. 

Mesos pools all the available CPUs, RAM, and disk space in a
cluster and provides an abstraction layer so that applications can
request arbitrary units of each resource. Mesos uses policies, such as
fair-sharing, to decide how to divide resources between co-scheduled
applications. Mesos uses {\it cgroups} to ensure isolation between
processes and also to guarantee that a process does not exceed its
allocation.

Another trend in the Big Data and Cloud space is that of {\it
containerization}, with Docker\cite{Merkel2014Docker:Deployment} emerging as a standard container choice. Containers provide a consistent development environment,
uniform packaging, ease of deployment, and a dynamic delivery mechanism
for new upgrades and new features. The scientific community has
embraced containers, as is evident from the recent availability of
several Dockerized scientific applications \cite{Ermakov2017TestingApplications} 
\cite{Morris2017UseSoftware}. 

While Apache Mesos recently added support for running Docker
containers, it does not have support for orchestration of Dockerized
scientific jobs. 
Native support for networking between Docker
containers is not available, which is essential for MPI based
scientific workloads that can span several VMs/nodes. To address this
shortcoming, we use Docker Swarm\cite{SwarmDocumentation}. The key advantage of Docker Swarm is how it provides cluster creation and container
scheduling. The Swarm nodes are connected to each other via a private
virtual network.

In this paper, we present the design and performance evaluation of our framework, {\it Scylla}, which uses Docker Swarm for communication between containerized tasks (MPI processes) and Apache Mesos as a resource manager. Scylla enables co-scheduling of multiple containers on a node, as well as the distribution of the containers in the cluster.

By default, the Docker Swarm manager decides which node will host the next container based on the Swarm's policy but that does not depend on the resource requirement of the application. Scylla allows a policy driven approach to determine which participating nodes will be hosting the application containers based on the resource requirement for each parallel application.

The contributions of this paper are the following:

\begin{itemize}

\item We present the design of our framework, Scylla, and the
architectural details that allow it to function as a Mesos framework
and also leverage Docker Swarm for orchestration of containerized
scientific jobs.

\item We present a policy driven approach to manage CPU, memory, and
networking resources more efficiently across VMs allocated on the
Chameleon cloud.

\item We present an evaluation of Scylla on a cluster of VMs on the Chameleon cloud and conclude on the efficacy of various policies for distributing containers with MPI processes.

\end{itemize}

\section{Background}
Apache Mesos consists of three major components (a) Mesos Agent  (b) Mesos Master and (c) the Mesos Framework. Mesos Agents are the computational nodes that execute the requested tasks. The Mesos Master acts as a resource broker between the agent and Mesos Frameworks for resource negotiation. After the negotiation, Mesos Framework makes scheduling decisions to map tasks to available agents. 
Each Mesos framework can either employ its own custom executor or use the default Mesos executor. The Executor takes the responsibility of executing a job within the Mesos defined environment.

\begin{figure}[h!]
  \centering
  	\includegraphics[width=0.5\textwidth]{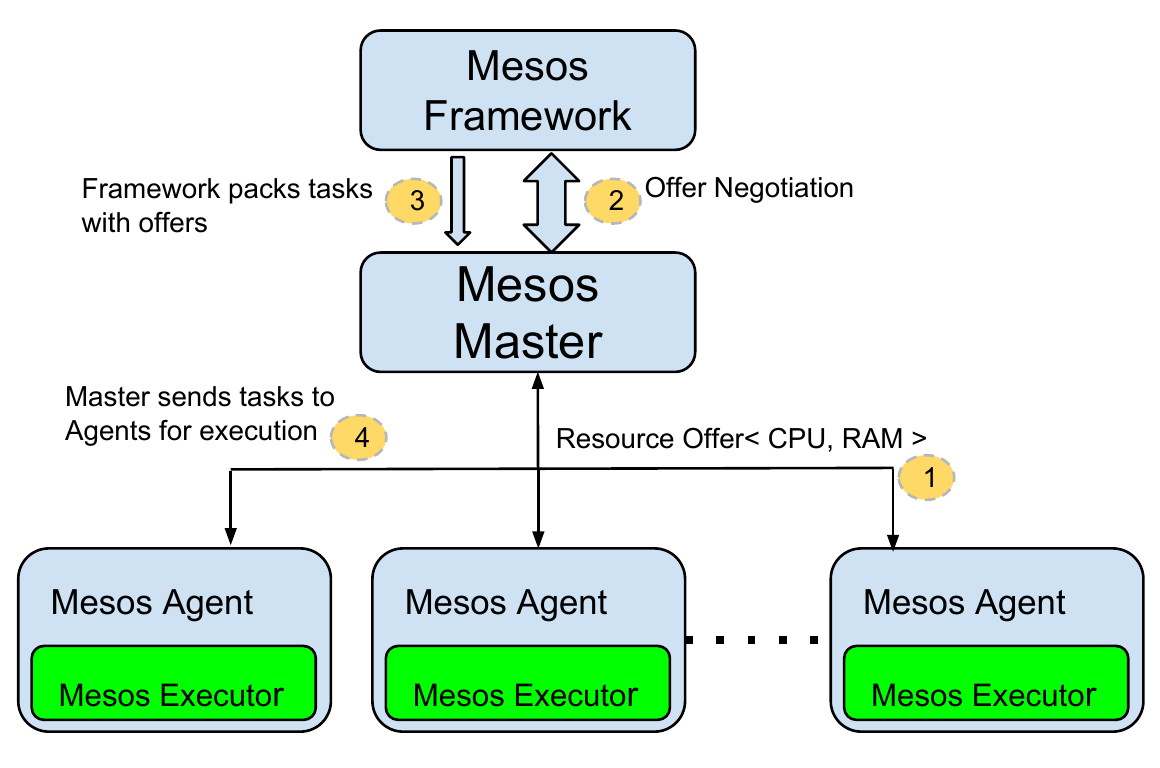}
      	\caption{{\it Apache Mesos Architecture}}
  	\label{Apache_Mesos_Architecture}
\end{figure}

Figure \ref{Apache_Mesos_Architecture} shows the steps involved when using a Mesos environment.
In step 1, all Mesos Agents advertise their available resources (CPU, RAM, disk, I/O) to the Mesos Master. In our current implementation we only use the CPU and RAM specifications. Next, in step 2, each framework negotiates offers with the Mesos Master. Mesos uses the Dominant Resource Fairness (DRF) algorithm by default. So, while a framework may get all or most of the dominant resource it has requested (CPU, for example), the other resource requests (for RAM and disk, for example) may not be satisfied. So, the negotiation phase involves frameworks possibly declining a few offers. After the negotiation phase, the frameworks have access to a list of available offers. In step 3, each framework, based on its scheduling policy, packs the tasks to the accepted offers. The framework then communicates its scheduling decisions with the Mesos Master.
In step 4, the Mesos Master send tasks to assigned Agents to execute it by using the appropriate Mesos executor.

\subsection{Apache Mesos Executor}
Mesos executors are part of the Mesos framework. Each Mesos framework can decide which executor to use for executing submitted jobs. The executor receives a task and executes it in the environment provided by each Mesos Agent. The executor can communicate with other software tools, such as Docker, to launch tasks. The executor terminates the task when it is completed or when Docker notifies that it has been completed. 

\subsection{Apache Mesos Framework}
A Mesos Framework is an external entity that communicates between the end user and the Mesos Master. It negotiates with the Mesos Master for resource offers. Upon approval, it employs the framework's scheduling policy to distribute the tasks among the Agents. Frameworks are expected to be designed as customizable in order to allow application specific policies and scheduling decisions. For example, Marathon~\cite{Marathon:DC/OS}, Apache Aurora~\cite{ApacheAurora}, Chronos~\cite{Chronos:Mesos}, Hadoop Framework~\cite{WelcomeHadoop}, and Spark Framework~\cite{ApacheComputing} are the widely used Mesos frameworks, each differing in the scheduling policy and also the nature of jobs they run.

\subsection{Docker Swarm}
Docker is a containerization technology that provides an isolated Operating System environment for each job. Apache Mesos provides native support for Docker containers \cite{ApacheContainerizer}. Docker Swarm provides container orchestration across multiple connected host machines. A node is a participating Docker engine for the Swarm cluster. 
Swarm can launch multiple replicas of the same Docker container image as part of the same service across multiple host machines connected via a Swarm overlay network. The overlay network is a software defined subnet on top of the host network. It can span across multiple host/VMs and allows containers on this network to seamlessly communicate.

\section{Architecture of Scylla with Apache Mesos and Docker Swarm}

Our Scylla framework is designed for MPI applications that could be distributed across several nodes. It uses Mesos to leverage resource pooling and enable co-scheduling of applications. The customizable policy driven approach allows optimized packing of MPI processes, each of which is launched in a Docker container, on the Mesos Agent nodes. Docker Swarm provides networking support between the containers running on the Agent nodes. 
Mesos uses its own containers for running tasks in a sandbox environment but it does not support inter-container communication. Mesos supports Docker containers but only to run in a standalone mode. The Docker Swarm overlay network provides the key feature to connect all the service containers and facilitate seamless inter-container communication. As a result, the Docker containers can be placed in different physical nodes while remaining a part of the same virtual private network. 

The Scylla framework allows the end user to choose the policy to distribute the MPI processes (each in a container) across nodes. This provides an opportunity to develop and deploy customized policies for the target applications to increase both cluster utilization and throughput. Scylla provides two policies to launch service containers across host nodes. 
\begin{itemize}
\item { ~\it Spread} - distribute the service containers on all the available nodes.
\item {~\it MinHost} - distribute the service containers in as few available nodes as possible. In Section V, our experimental results show how these two policies can be useful to increase the cluster resource utilization.
\end{itemize}


The Scylla framework has a custom Mesos executor that communicates with Docker Swarm Manager to run the Mesos tasks across multiple Docker containers. 

The architectural setup of Scylla with a Mesos and Docker Swarm cluster consists of the following components.
\begin{itemize}
\item Node-0: this is a host node in the cluster that hosts the Mesos Master, Mesos Agent with custom Scylla executor and Swarm Manager. 
\item Task-0: this task is created by Scylla to initiate the launch of the MPI task in the service containers. Task-0 uses the custom Scylla executor to communicate with the Swarm Manager to orchestrate the containers for running the MPI task.
\item  Workers: these are the computational nodes that consists of Mesos Agents and Swarm Workers.
\end{itemize}

\begin{figure}[h!]
 \vspace{-1em}
  \includegraphics[width=0.5\textwidth]
  {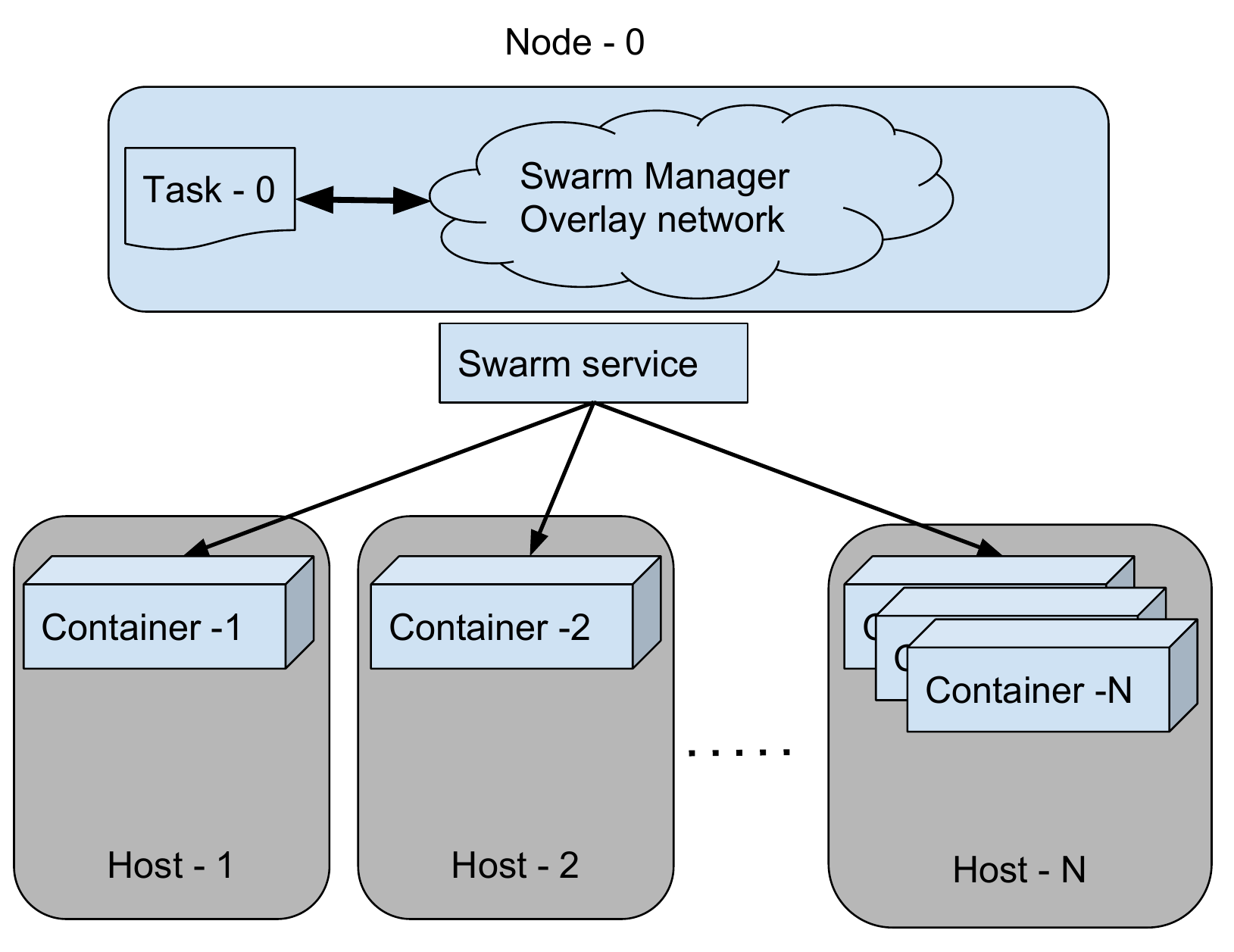}
  \caption{{\it Docker Swarm service containers spread across an overlay network}}
  \label{Docker_swarm_overlay_network}
\end{figure}

Figure \ref{Docker_swarm_overlay_network} is a representation of how Docker Swarm containers are placed in a distributed cluster. The Docker containers of the Swarm service are spread across multiple host nodes but connected to the same overlay network, which presents all containers in the same logical network.
 
 \begin{figure}[h!]  
  \centering
  \includegraphics[width=0.5\textwidth]
  {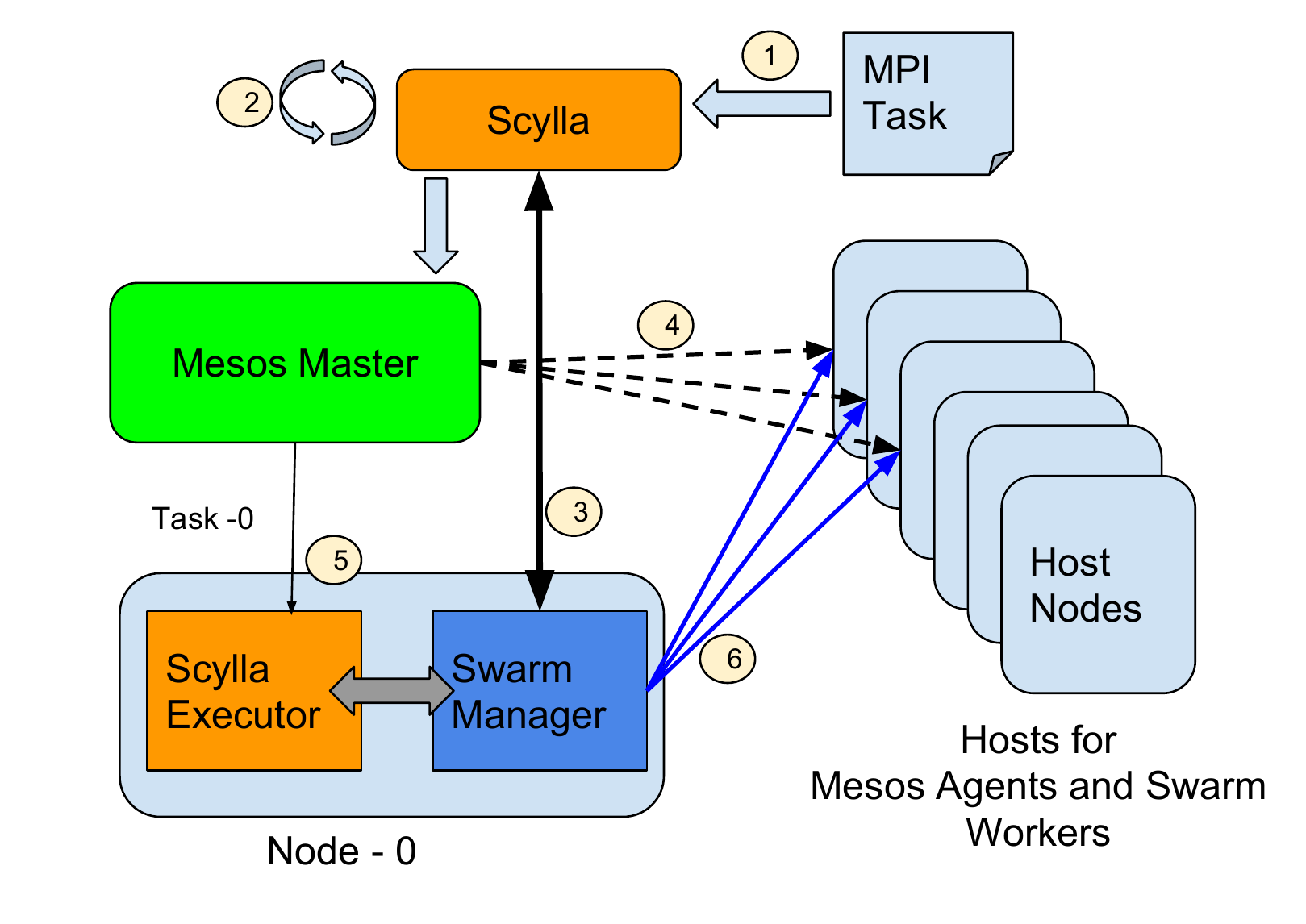}
  \caption{{\it Event flow of an MPI task on Mesos-Docker Swarm cluster through Scylla}}
  \vspace{-0.5em}
  \label{Event_flow_Mesos_Docker_Swarm_Scylla}
\end{figure}

Figure \ref{Event_flow_Mesos_Docker_Swarm_Scylla} illustrates the event flow when a Dockerized MPI task is launched via the Scylla framework. 
\begin{enumerate}
	\item User submits an MPI application by selecting an MPI Docker image from Docker Hub\cite{DockerHub}. Scylla has a web interface that provides options to select and specify the parameters for the application.
    
    \item Scylla, executing on the Node-0, negotiates the offers with the Mesos Master and receives available offers.
    
    \item Based on the requirements of the MPI application and the policy of the framework, Scylla chooses appropriate offers and creates the service containers on the chosen agent nodes. It also keeps track of how the containers are distributed on the agent nodes. One of the containers it launches is the Master container that creates an {\it SSH} tunnel to the head node. For each MPI application, it creates an SSH tunnel between the master container and the head node. Once all the service containers are created successfully, Scylla collects their IP addresses on the overlay network and sets the {\it hostfile} inside the master container.
         
    \item Scylla launches Mesos tasks to block the amount of allocated resources for the service containers in each of the participating nodes. 
    
    \item Once all the tasks are launched successfully, Scylla launches an {\it init} task, named task-0, on Node-0 through custom Scylla executor that communicates with the Swarm Manager and connects with the master service container through the SSH tunnel.
    
    \item Task-0 starts the MPI application processes via the master service container, which distributes the processes across the worker containers in the Mesos Agents.
    
\end{enumerate}
 
\section{Experimental Setup}

\subsection{MPI Benchmarks}
We chose a set of Dockerized MPI applications that represent scientific workloads and computational kernels. They are described in Table 1. To Dockerize these applications we have used {\it nlknguyen/alpine-mpich:onbuild} \cite{Nlknguyen/alpine-mpichHub} as our base image. It contains MPICH~\cite{MPICHMPI} and other required softwares to enable MPI programs to execute. We have created individual Docker images for each application along with all the dependencies.  

\begin{table}[h!]
\centering
\renewcommand{\arraystretch}{1.5}
\begin{tabular}{|l|p{0.70\linewidth}|}
 \hline
 {\bf Benchmark} & {\bf Description}\\
 \hline
 CoMD\cite{Heroux2009ImprovingMini-applications} 		& Classical molecular dynamics algorithms\\ 
 \hline
 MiniAMR\cite{Heroux2009ImprovingMini-applications} 	& Adaptive mesh refinement\\
 \hline
 MiniFE\cite{Heroux2009ImprovingMini-applications} 	& Unstructured finite element solver\\
 \hline
 HPCCG \cite{Heroux2009ImprovingMini-applications}		& A conjugate gradient solver\\
 \hline
 MiniAero\cite{Heroux2009ImprovingMini-applications} 	& Unstructured finite volume code\\
 \hline
 HP2P\cite{CEADevelopment} 	    & Measures bandwidths and the latencies of a network\\
 \hline
\end{tabular}
\smallskip
\label{Mantevo_Mini_Application_Suite}
\caption{\textit{ MPI Mini-Applications }}
\end{table} 
\vspace{-1em}
\begin{table}[h!]
\centering
\renewcommand{\arraystretch}{1.5}
\begin{tabular}{|l|p{0.70\linewidth}|}
 \hline
 {\bf Software} & {\bf Version}\\
 \hline
 Ubuntu 		& Ubuntu 16.04.2 LTS (Xenial)\\ 
 \hline
 Docker Engine 	& 17.06.0-ce\\
 \hline
 Apache Mesos 	& 1.3.0\\
 \hline
 
\end{tabular}
\smallskip
\label{Software Stack and Vession}
\caption{\textit{Software Stack and Version}}
\end{table} 









\begin{figure}[h!] 
  \centering
  \includegraphics[width=0.5\textwidth]
  {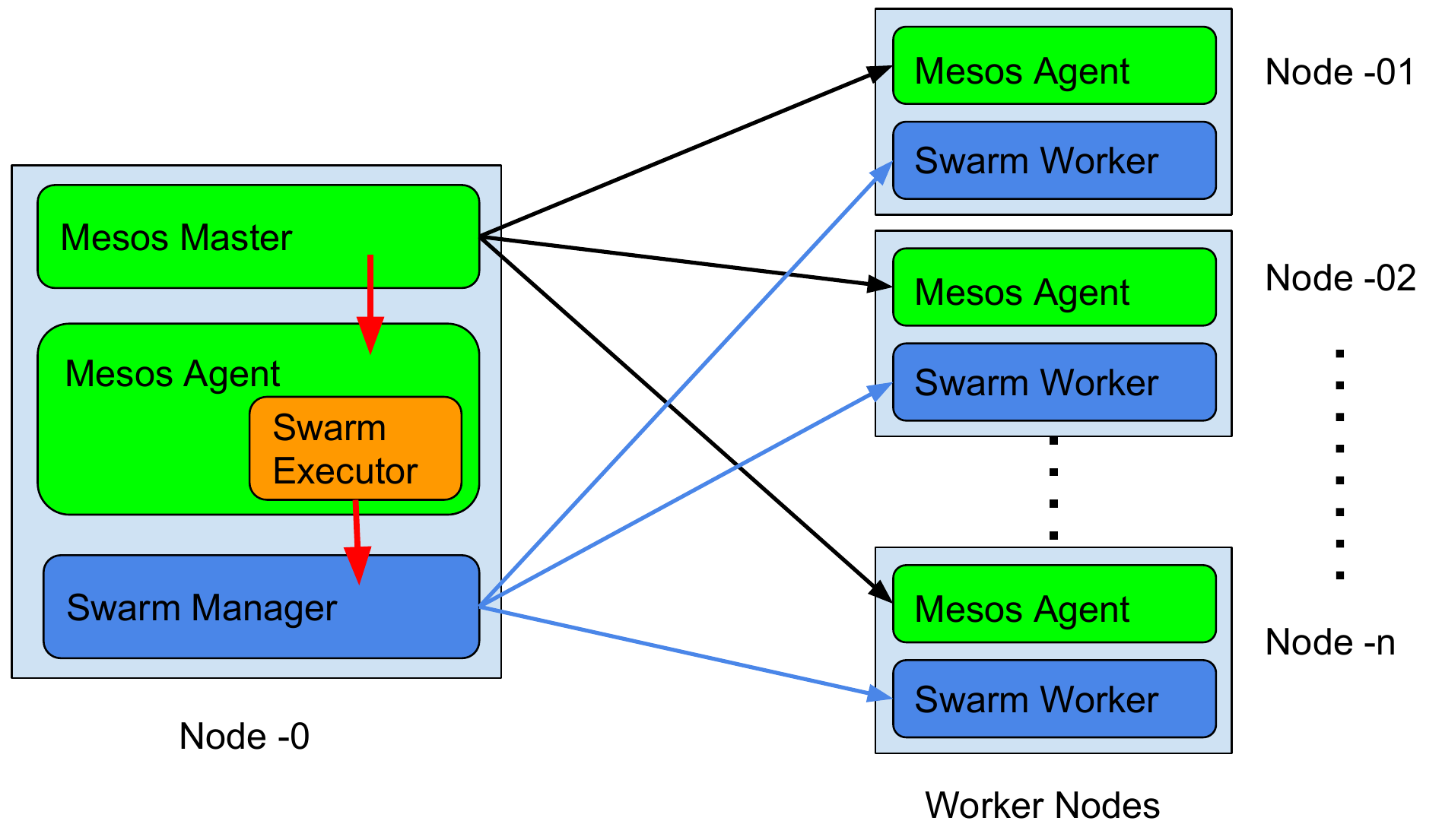}
  \caption{\textit{Experimental setup of an overlay cluster of Apache Mesos and Docker Swarm}}
  \vspace{-1.5em}
  \label{mesos_swarm_node_setup}
\end{figure}

\subsection {Experimental Cluster}
Figure~\ref{mesos_swarm_node_setup} shows our experimental cluster on the Chameleon cloud. Each node contains 48 CPU cores and 124GB of RAM. This allocation is typical of what scientists are given on the Chameleon cluster. 

We also received another KVM based Virtual Machine (VM) node with smaller resource configuration to serve as Node-0 (head node). Node-0 hosts the Mesos Master, Swarm Manager, Mesos Agent, along with a custom Scylla MPI-Swarm executor. Each of the bare metal nodes contain a Mesos Agent advertising its resources to the Mesos Master and a Swarm Worker that commits the resources to the Swarm Manager. Resource offers from the Mesos Agent at Node-0 is not considered for any submitted tasks and the Swarm Manager is set up to not run any service containers. 


After the resource negotiation phase, Scylla decides which Mesos Agents are the best fit to execute the application. Scylla launches Task-0 on Node-0 that communicates with the underlying Swarm Manager to launch Dockerized MPI task as a service on the specific nodes identified by Scylla. Mesos works as the resource manager whereas the Docker Swarm Manager is the container orchestrator. 
\section{Experimental Results}
\subsection{Overhead of MPI Processes within Containers}

\begin{figure}[h!]
  \includegraphics[width=0.5\textwidth]
  {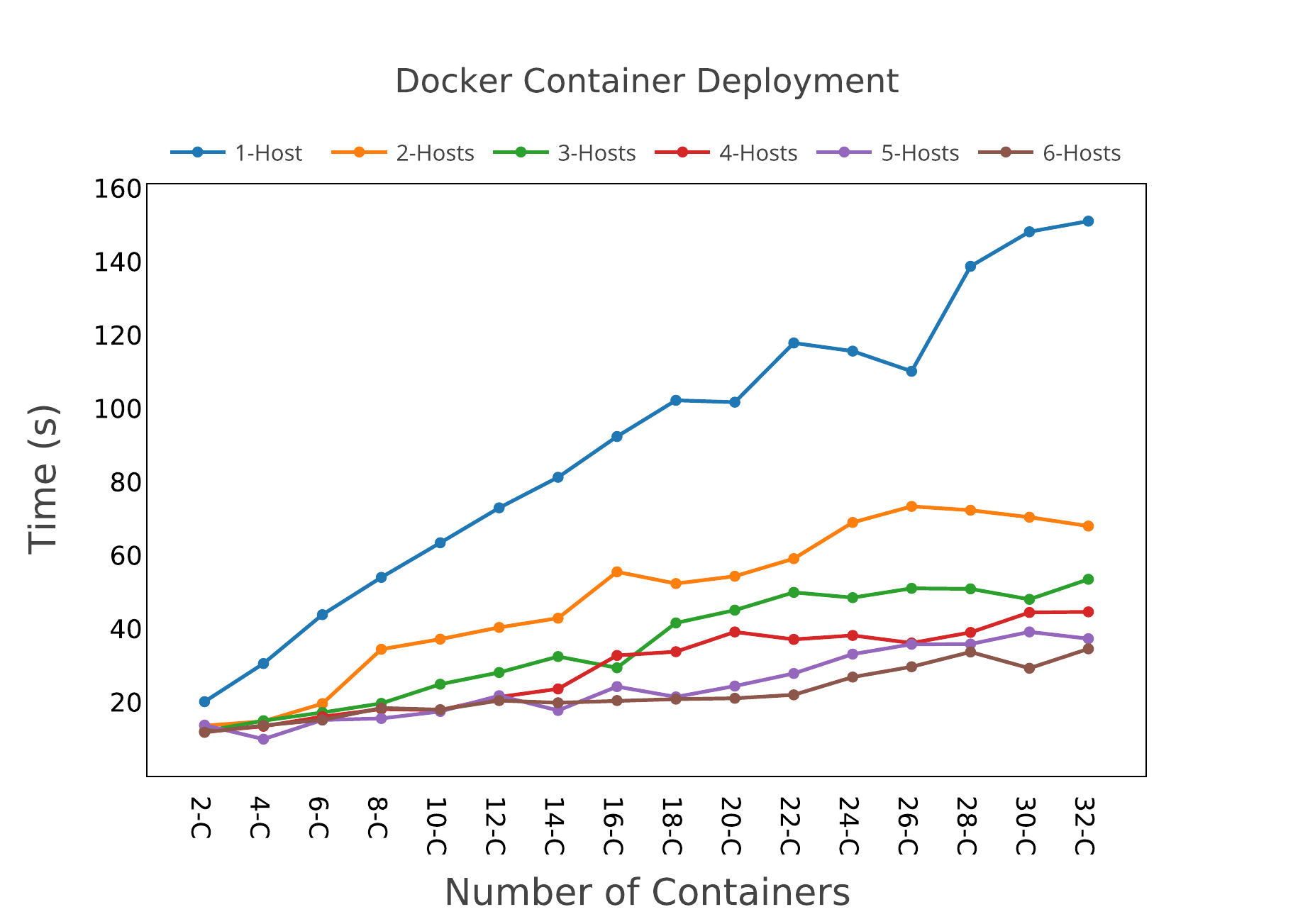}
    \caption{{\it Container deployment overhead}}
    \vspace{-1em}
  \label{Container_deployment_overhead} 
\end{figure}

Scylla creates one service container for each parallel process of an MPI task. Based on the MPI application, increased parallelism may improve the performance, but in the Scylla implementation, increased parallelism has an additional overhead for creating more containers. Container creation cost varies depending the size of the cluster. We have varied the number of hosts in our experimental cluster and observed the container overhead. 

In Figure~\ref{Container_deployment_overhead}, we observe that increasing the number of hosts in a cluster reduces the overhead of creating the required service containers. We varied the cluster size from 2 to 6 hosts and the result shows that for a cluster size greater than 3 hosts, where less than 16 containers are deployed, the overhead does not vary significantly. 
One service container is associated with each MPI parallel process. Our results show that if the required parallelism is below 16, for a cluster of size 4 or greater, the service container creation overhead is close to 20\% compared to the overall run time. This can be attributed to the nature of our workloads that consist of mini-scale MPI applications. We expect this overhead to be much lower for long running MPI applications. So, the container overhead is an important consideration when the required parallelism is more than 16 and the size of the cluster is below 4 nodes in our experimental setup.

\begin{figure}[h!]
\vspace{-1em}
  \includegraphics[width=0.5\textwidth]  {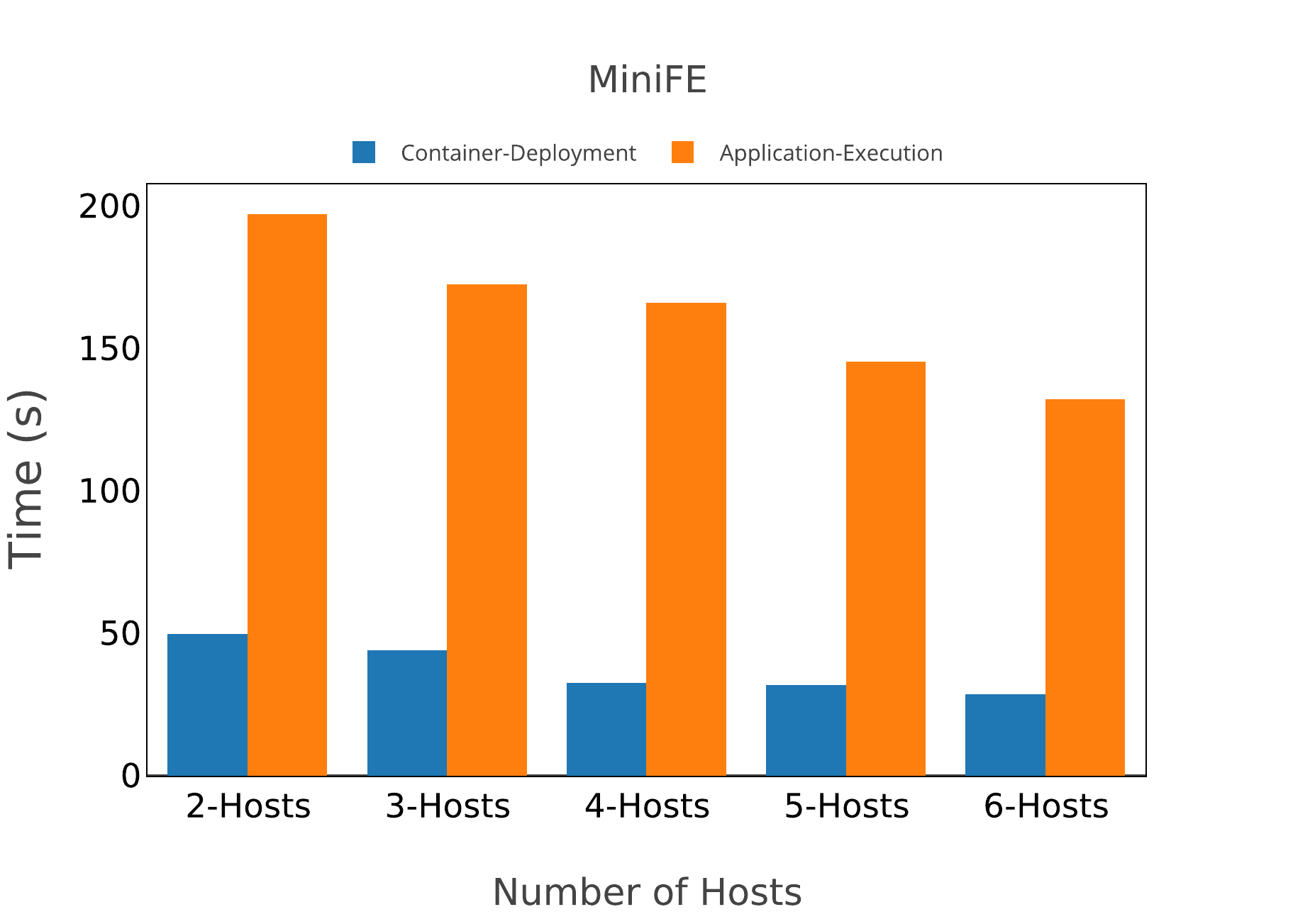}
   \caption{{\it MiniFE with variable cluster size}}
  \label{MiniFE_with_variable_cluster_size}
\end{figure}

In Figure \ref{MiniFE_with_variable_cluster_size}, we  show how the execution time of the MiniFE MPI application, which is both CPU and memory intensive, increases as we vary the number of nodes in the cluster. As the container instantiation overhead cost is amortized due to the execution time of MiniFE, it is expectedly beneficial to distribute it to as many nodes as available. 
\vspace{-1em}
\begin{figure}[h!]
  \includegraphics[width=0.5\textwidth]
  {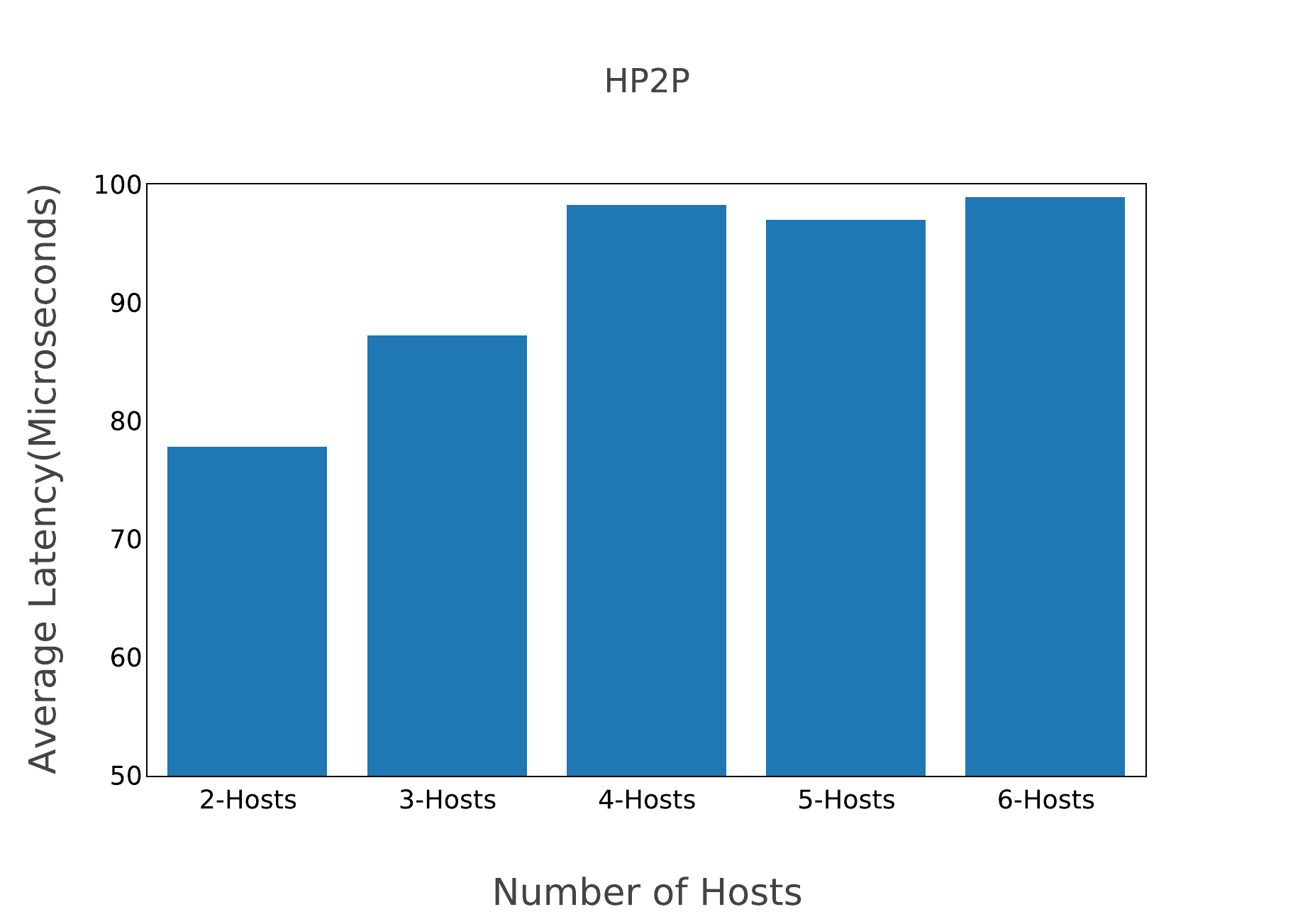}
    \caption{{\it Communication intensive HP2P Benchmark}}
    \vspace{-1em}
   \label{Communication_Heavy_Workload}
\end{figure}

\vspace{3mm}
Figure~\ref{Communication_Heavy_Workload} demonstrates the performance of the MPI benchmark, HP2P, which communicates with all the hosts and computes the average latency for information sent over the Docker Swarm network. We ran HP2P for 32 parallel Docker occurrences with 2048 MB of data to be transmitted for 20 iteration and varied the number of hosts in the cluster. We observe that the required transmission time increases as we increase the size of the cluster. The average latency increases by 10\% till the size of the cluster reaches four. After that, as nodes are added to the cluster, there is no significant change. These results do not change even when we ran the experiment with 64 Docker instances and 4096 MB of information is sent across the network for 20 iterations.

\subsection{Co-Scheduling MPI Tasks Across Host Nodes}
In the approach used often in HPC clusters, MPI jobs get exclusive access to the nodes. While this traditional approach avoids container overheads, it can suffer from low resource utilization and throughput compared to the co-scheduled approach.
\vspace{-1em}
\begin{figure}[h!]
  \includegraphics[width=0.5\textwidth]
  {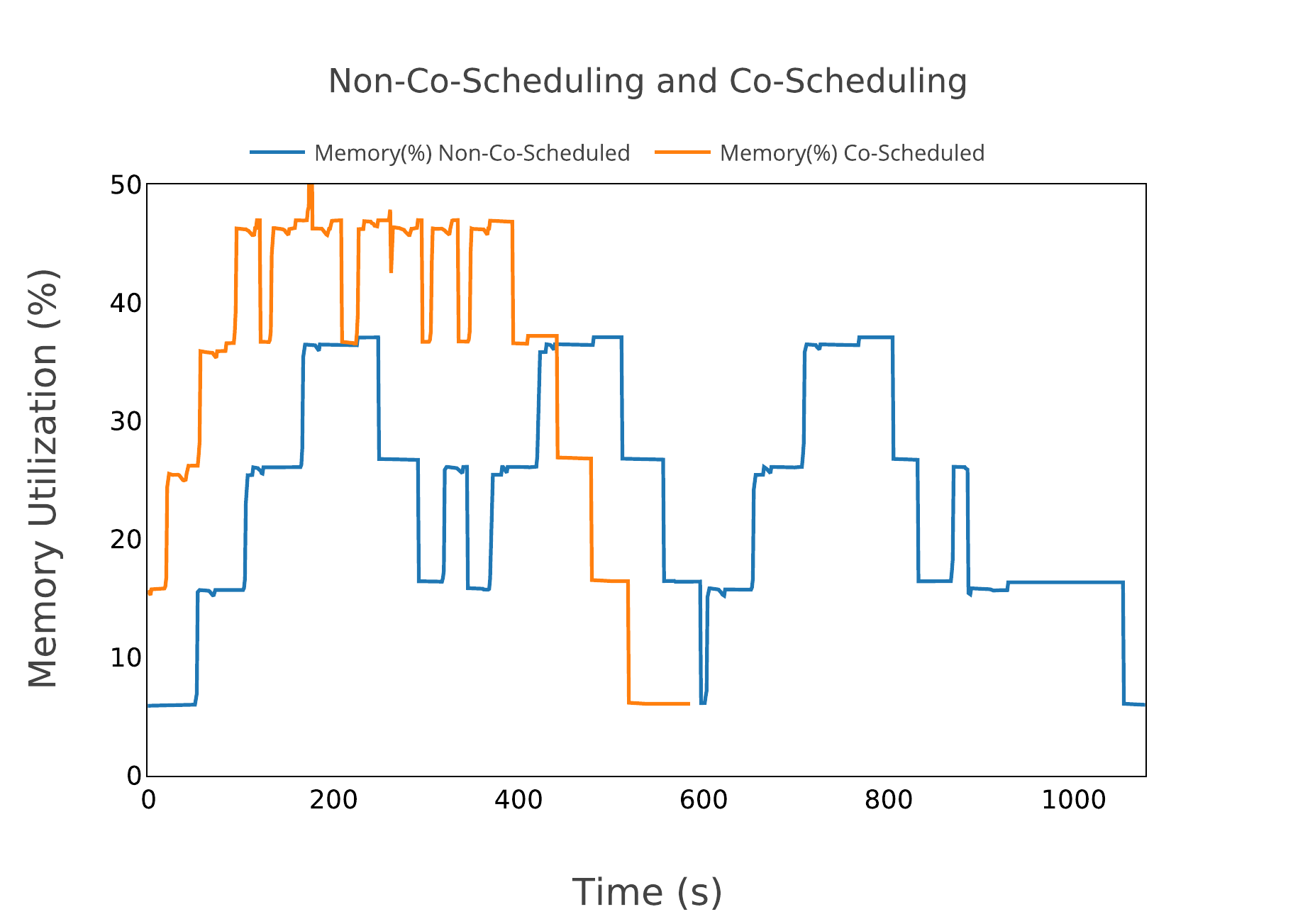}
    \caption{{\it Memory Utilization in a Non-Co-Scheduled vs. Co-Schedule Approach}}
    \vspace{-1.5em}
    \label{memory-scheduling-policies}
\end{figure}

\begin{figure}[h!]  
  \includegraphics[width=0.5\textwidth]
  {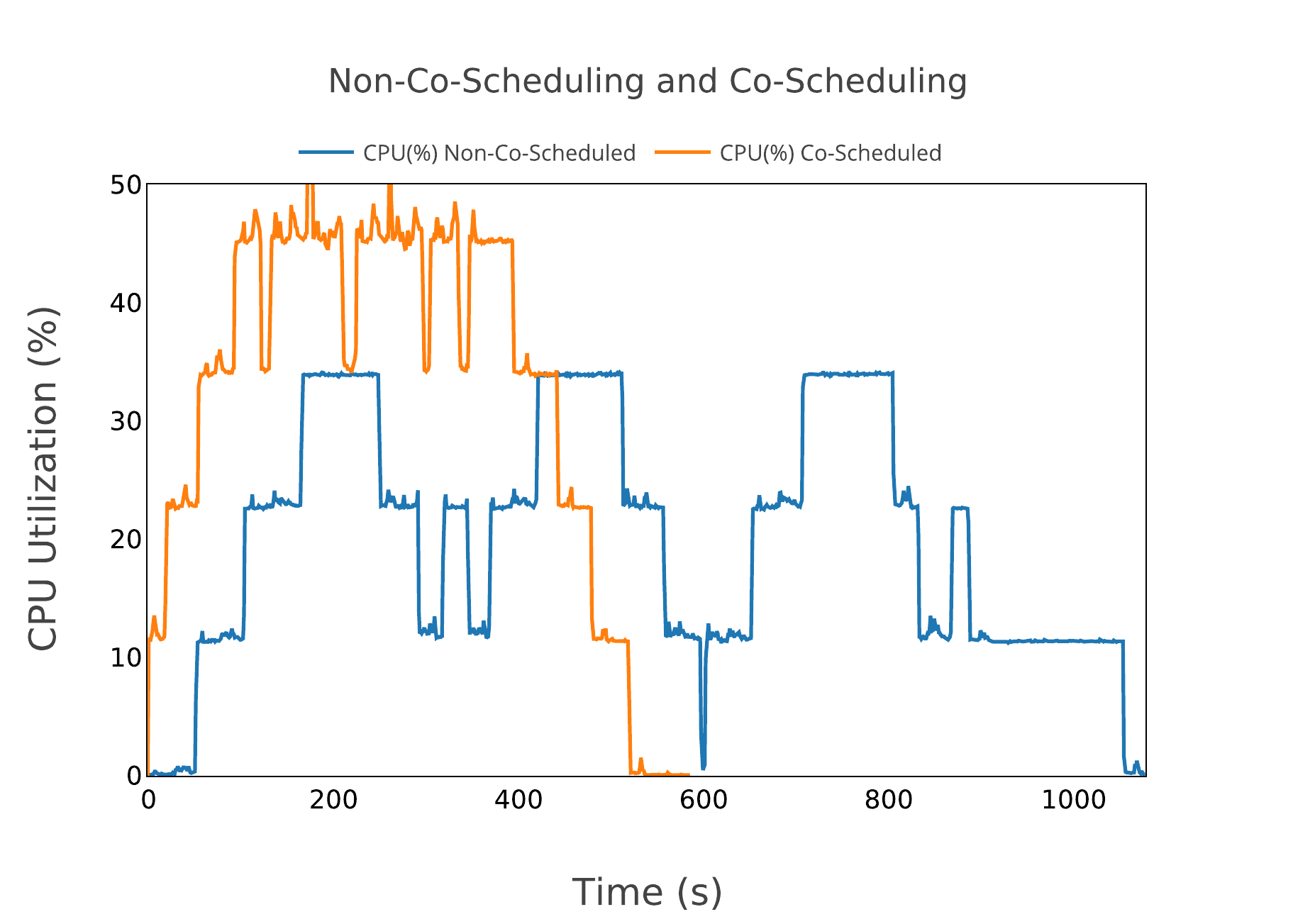}
  \caption{{\it CPU Utilization in Non-Co-Scheduled vs. Co-Scheduled Approach}}
    \vspace{-1.5em}
    \label{cpu-scheduling-policies}
\end{figure}
\vspace{1em}
Figure~\ref{memory-scheduling-policies} and~\ref{cpu-scheduling-policies} demonstrate CPU and memory usage using the traditional HPC (without co-scheduling) and the Mesos co-scheduling approach. We ran ten occurrences of the MiniFe MPI application using both the approaches and gathered performance data using the Performance Co-Pilot software tool \cite{PerformanceCo-Pilot}. The performance data quantifies the gains in resource utilization using the co-scheduled approach. Without co-scheduling, the CPU usage peaks around 35\% and  dips even further several times. In the co-scheduled approach, it peaks at 45\% but stays around the peak longer than in the non co-scheduled approach. 
For the same number of tasks, the co-scheduled approach takes close to half of the time to finish those tasks compared to that of the non-co-scheduled approach.
\vspace{-1.5em}
\begin{figure}[h!]  
  \includegraphics[width=0.5\textwidth]
  {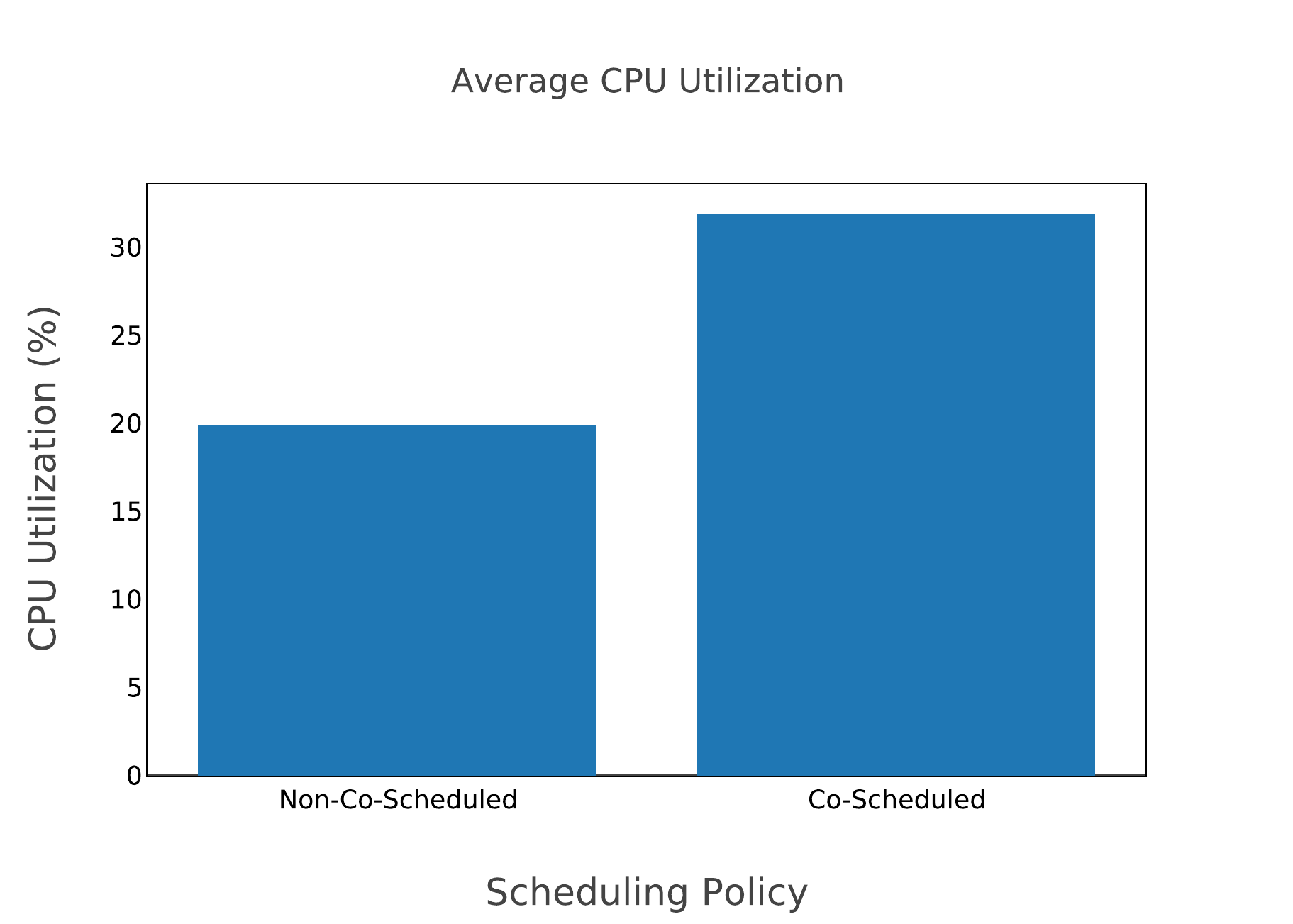}
  \caption{{\it Average CPU Utilization in Non-Co-Scheduled vs. Co-Scheduled Approach}}
  \vspace{-1.5em}
    \label{cpu-scheduling-policies-bar}
\end{figure}

\begin{figure}[h!]  
  \includegraphics[width=0.5\textwidth]
  {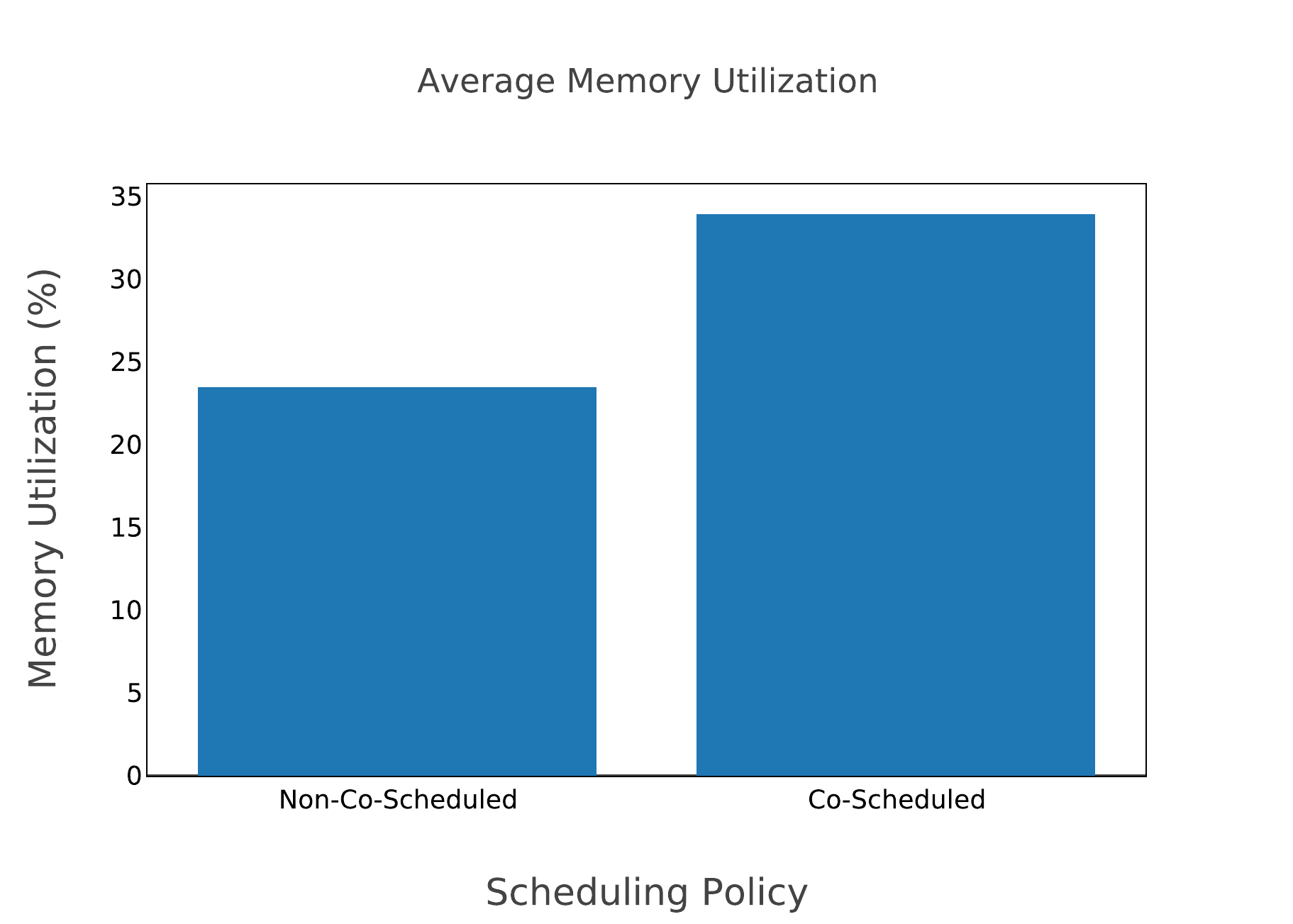}
  \caption{{\it Average Memory Utilization in Non-Co-Scheduled vs. Co-Scheduled Approach}}
   \vspace{-1.5em}
    \label{mem-scheduling-policies-bar}
\end{figure}
 \vspace{1em}
Figure \ref{cpu-scheduling-policies-bar} and \ref{mem-scheduling-policies-bar} present an average of the CPU and memory utilization for non-Co-Scheduled and Co-Scheduled Policy.
We have observed that CPU utilization is 60\% and memory utilization is 44\% more in Scylla-imposed co-scheduled approach compared to the non-Co-Scheduled approach.


\subsection{Policy Driven Container Placement in Docker Swarm}

Based on these observations of the experimental results, it is clear that MPI applications need to be launched in a Swarm cluster with different policies for container placement based on the type of the application. {\it Scylla}, provides two policies:
\begin{itemize}
\item {\it Spread} - Scylla finds all eligible offers from the Mesos Master and distributes the service containers across all of them. This policy is enforced by Scylla for resource intensive tasks in order to reduce the resource contention.
\item {\it MinHost} - In this approach, Scylla finds all eligible offers and picks the minimum number of Mesos Agents that can fit all the required service containers with specific resource requirements. The goal of this policy is to keep the containers as co-located as possible so that the overhead due to network communication can be reduced. This policy is enforced by Scylla for communication intensive tasks.  
\end{itemize}
\vspace{-1em}
\begin{figure}[h!] 
  \includegraphics[width=0.5\textwidth]
  {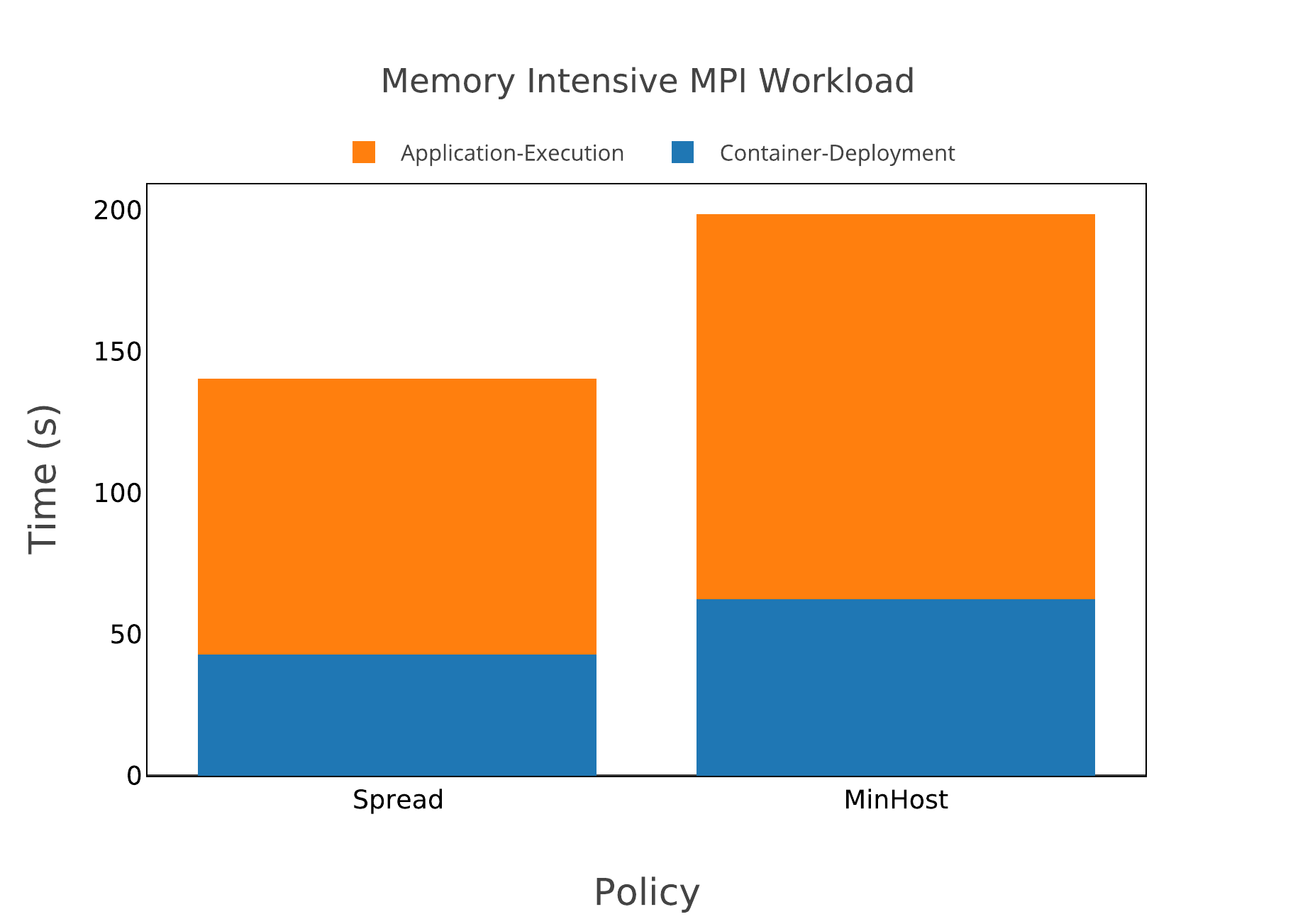}
  \caption{{\it Memory Intensive Workload Applied Policies}}
  \vspace{-1.5em}
    \label{Resource_Heavy_Workload_Different_Policies}
\end{figure}

\begin{figure}[h!]
  \includegraphics[width=0.5\textwidth]
  {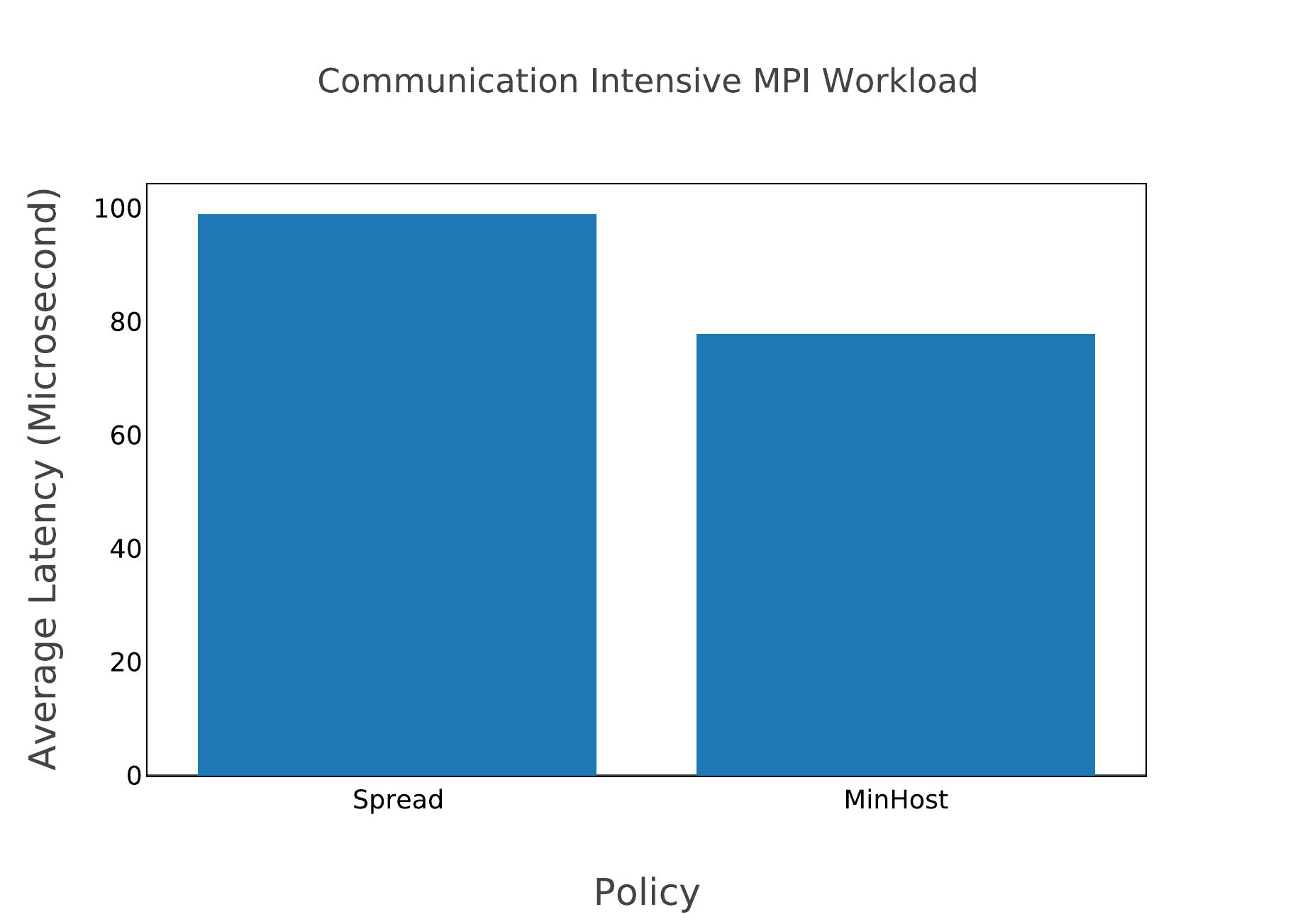}
  \caption{{\it Communication Intensive Workload Applied Policies}}
   \vspace{-1em}
  \label{Communication_Heavy_Workload_Different_Policies}
\end{figure}
\vspace{1.5mm}
We ran the memory intensive MPI application, MiniFE, and the communication intensive benchmark, HP2P, to measure the performance of both the policies.
Our experimental results in Figure \ref{Resource_Heavy_Workload_Different_Policies} and Figure \ref{Communication_Heavy_Workload_Different_Policies} show that Spread is better for resource intensive tasks whereas the MinHost policy is better for a communication intensive task. miniFE perform 29\% better with the Spread  policy while HP2P shows 21\% better average latency using the MinHost policy compared to Spread.


\section{Related Work}

Sparks~\cite{Sparks2017HPCUse}  highlighted  that attributes such as host level access, privileged operations, and runtime environments are essential to run a Docker container in any environment. For example, Singularity~\cite{Kurtzer2017Singularity:Compute}, a containerization application, mitigates two of the main bottlenecks faced by administrators in the HPC community -- unauthorized user root privileges and portability. Singularity makes an important contribution to the container space, but it does not currently have orchestration support which Scylla aims to provide. 

In our earlier work~\cite{Saha2016IntegratingMesos}, we discussed the feasibility of enabling Docker containers based scientific application to run va Science Gateways\cite{HomeGateways}. In this current work, we have presented a design and implemented a framework which can schedule containerized MPI tasks on HPC clusters.

Abdelbaky et al.~\cite{Abdelbaky2015DockerCenters} have shown an approach to launch Docker containers across multiple clouds and addressed the challenge of container scheduling and placement. Our work places the container of a single service across multiple host machines  wherein the scheduler is aware of the host machines and their resources. We utilize the resources of each host based on the submitted request and organize the placement of containers across hosts to improve overall execution time.

Nguyen et al.~\cite{Nguyen2017DistributedMode} show how Docker Swarm can be utilized to create an environment for running MPI applications on the cloud. We have extended that work by providing (1) integration with Apache Mesos; (2) resource matching for each MPI task; and (3) a policy driven framework that decides how to distribute MPI tasks and related service containers across host nodes. 


Ruiz C et al.~\cite{Ruiz2015PerformanceHPC} performed an evaluation of container performance for HPC environments. In the experimental setup, they ran NAS benchmarks on containers located on single physical machines and also spread across several physical machines. They observed that the inter-container communication is fast but there is a degradation of CPU performance for memory intensive applications due to shared memory communication. They also noted that the virtual network for inter-container communication does not add any extra cost. However when containers were distributed to several physical nodes, they observed that network bound applications can be affected severely by the container network. Our results in Figure~\ref{MiniFE_with_variable_cluster_size} and \ref{Communication_Heavy_Workload} show similar trends for memory intensive and network communication intensive MPI applications that we ran.

It is to be noted that several research initiatives have done extensive work on containers for HPC environments. Bahls~\cite{Bahls2016EvaluatingApplications} conducted experiments using Cray XC40™ systems~\cite{CraySystems}, in which the author was able to identify the potential of containers for HPC to run serial as well as parallel applications. This work is not akin to the research presented in our research, but it is a contribution to the HPC community that serves as an incentive for full container technology adoption in HPC. 

\section{Conclusion}


{\it Scylla} provides fine grained resource allocation, through Apache Mesos, for the task launched in a cluster and distributes the MPI task according to the nature of the application to improve the overall throughput.

Expectedly, inter-container communication increases the over head when the containers are spread across multiple physical hosts. So, for a network intensive workload, {\it Scylla} provides a policy to keep the containers close to each other. On the other hand, for CPU and memory intensive tasks, keeping the containers close, increases the resource contention and thereby decreases the overall performance. Scylla's policy driven scheduling distributes the containers based on the nature of the task and mitigates the overhead due to resource contention and virtual network communication across physical hosts. Apart from container distribution, Scylla enables and enforces co-scheduling through the Apache Mesos resource manger, which improves both the resource utilization and throughput.

\section{Future Work}
Containers, such as the ones provided by Docker, have emerged as a widely acceptable solution for easy application deployment. However, there are other new technologies that are competing with Docker to provide specialized features for application development and deployment.  Singularity~\cite{Kurtzer2017Singularity:Compute} and rkt~\cite{RktCoreOS} can also support the MPI-HPC paradigm with container orchestrators such as Kubernetes\cite{KubernetesOrchestration}. The Docker Swarm mode enables communication between containers via a private overlay network. For the other container mechanisms, overlay network or Network Address Translation (NAT) modules are not currently available. We plan to study these technologies once these features are available. We also plan to design solutions with resource managers such as Apache Hadoop Yarn \cite{Vavilapalli2013ApacheYARN} and frameworks for Apache Spark\cite{Zaharia2016ApacheSpark} to allow the HPC community to adopt more cloud technologies into the HPC ecosystem.

\bibliographystyle{IEEEtran}
\bibliography{main.bib}

\begin{thebibliography}{10}
\providecommand{\url}[1]{#1}
\csname url@samestyle\endcsname
\providecommand{\newblock}{\relax}
\providecommand{\bibinfo}[2]{#2}
\providecommand{\BIBentrySTDinterwordspacing}{\spaceskip=0pt\relax}
\providecommand{\BIBentryALTinterwordstretchfactor}{4}
\providecommand{\BIBentryALTinterwordspacing}{\spaceskip=\fontdimen2\font plus
\BIBentryALTinterwordstretchfactor\fontdimen3\font minus
  \fontdimen4\font\relax}
\providecommand{\BIBforeignlanguage}[2]{{%
\expandafter\ifx\csname l@#1\endcsname\relax
\typeout{** WARNING: IEEEtran.bst: No hyphenation pattern has been}%
\typeout{** loaded for the language `#1'. Using the pattern for}%
\typeout{** the default language instead.}%
\else
\language=\csname l@#1\endcsname
\fi
#2}}
\providecommand{\BIBdecl}{\relax}
\BIBdecl

\bibitem{ChameleonCloud}
\BIBentryALTinterwordspacing
``{Chameleon Cloud}.'' [Online]. Available:
  \url{https://www.chameleoncloud.org/}
\BIBentrySTDinterwordspacing

\bibitem{Jetstream:Cloud}
\BIBentryALTinterwordspacing
``{Jetstream: A National Science and Engineering Cloud}.'' [Online]. Available:
  \url{https://jetstream-cloud.org/}
\BIBentrySTDinterwordspacing

\bibitem{Stewart2015Jetstream}
\BIBentryALTinterwordspacing
C.~A. Stewart, G.~Turner, M.~Vaughn, N.~I. Gaffney, T.~M. Cockerill, I.~Foster,
  D.~Hancock, N.~Merchant, E.~Skidmore, D.~Stanzione, J.~Taylor, and S.~Tuecke,
  ``{Jetstream},'' in \emph{Proceedings of the 2015 XSEDE Conference on
  Scientific Advancements Enabled by Enhanced Cyberinfrastructure - XSEDE
  '15}.\hskip 1em plus 0.5em minus 0.4em\relax New York, New York, USA: ACM
  Press, 2015, pp. 1--8. [Online]. Available:
  \url{http://dl.acm.org/citation.cfm?doid=2792745.2792774}
\BIBentrySTDinterwordspacing

\bibitem{SlurmManager}
\BIBentryALTinterwordspacing
``{Slurm Workload Manager}.'' [Online]. Available:
  \url{https://slurm.schedmd.com/}
\BIBentrySTDinterwordspacing

\bibitem{TORQUEManager}
\BIBentryALTinterwordspacing
``{TORQUE Resource Manager}.'' [Online]. Available:
  \url{http://www.adaptivecomputing.com/products/open-source/torque/}
\BIBentrySTDinterwordspacing

\bibitem{Hindman2011Mesos:Center}
\BIBentryALTinterwordspacing
B.~Hindman, A.~Konwinski, M.~Zaharia, A.~Ghodsi, A.~D. Joseph, R.~Katz,
  S.~Shenker, and I.~Stoica, ``{Mesos: a platform for fine-grained resource
  sharing in the data center},'' pp. 295--308, 2011. [Online]. Available:
  \url{http://dl.acm.org/citation.cfm?id=1972488}
\BIBentrySTDinterwordspacing

\bibitem{MesosNews}
\BIBentryALTinterwordspacing
``{Mesos has been simulated to scale to 50,000 nodes, although it is not clear
  ho... | Hacker News}.'' [Online]. Available:
  \url{https://news.ycombinator.com/item?id=10228820}
\BIBentrySTDinterwordspacing

\bibitem{Merkel2014Docker:Deployment}
\BIBentryALTinterwordspacing
D.~Merkel, ``{Docker: Lightweight Linux Containers for Consistent Development
  and Deployment},'' \emph{Linux J.}, vol. 2014, no. 239, 3 2014. [Online].
  Available: \url{http://dl.acm.org/citation.cfm?id=2600239.2600241}
\BIBentrySTDinterwordspacing

\bibitem{Ermakov2017TestingApplications}
\BIBentryALTinterwordspacing
A.~Ermakov and A.~Vasyukov, ``{Testing Docker Performance for HPC
  Applications},'' \emph{CoRR}, vol. abs/1704.05592, 2017. [Online]. Available:
  \url{http://arxiv.org/abs/1704.05592}
\BIBentrySTDinterwordspacing

\bibitem{Morris2017UseSoftware}
\BIBentryALTinterwordspacing
D.~Morris, S.~Voutsinas, N.~C. Hambly, and R.~G. Mann, ``{Use of Docker for
  deployment and testing of astronomy software},'' \emph{CoRR}, vol.
  abs/1707.03341, 2017. [Online]. Available:
  \url{http://arxiv.org/abs/1707.03341}
\BIBentrySTDinterwordspacing

\bibitem{SwarmDocumentation}
\BIBentryALTinterwordspacing
``{swarm mode | Docker Documentation}.'' [Online]. Available:
  \url{https://docs.docker.com/engine/swarm/swarm-tutorial/}
\BIBentrySTDinterwordspacing

\bibitem{Marathon:DC/OS}
\BIBentryALTinterwordspacing
``{Marathon: A container orchestration platform for Mesos and DC/OS}.''
  [Online]. Available: \url{https://mesosphere.github.io/marathon/}
\BIBentrySTDinterwordspacing

\bibitem{ApacheAurora}
\BIBentryALTinterwordspacing
``{Apache Aurora}.'' [Online]. Available: \url{http://aurora.apache.org/}
\BIBentrySTDinterwordspacing

\bibitem{Chronos:Mesos}
\BIBentryALTinterwordspacing
``{Chronos: Fault tolerant job scheduler for Mesos}.'' [Online]. Available:
  \url{https://mesos.github.io/chronos/}
\BIBentrySTDinterwordspacing

\bibitem{WelcomeHadoop}
\BIBentryALTinterwordspacing
``{Welcome to Apache™ Hadoop{\textregistered}!}'' [Online]. Available:
  \url{http://hadoop.apache.org/}
\BIBentrySTDinterwordspacing

\bibitem{ApacheComputing}
\BIBentryALTinterwordspacing
``{Apache Spark™ - Lightning-Fast Cluster Computing}.'' [Online]. Available:
  \url{https://spark.apache.org/}
\BIBentrySTDinterwordspacing

\bibitem{ApacheContainerizer}
\BIBentryALTinterwordspacing
``{Apache Mesos - Supporting Container Images in Mesos Containerizer}.''
  [Online]. Available:
  \url{http://mesos.apache.org/documentation/latest/container-image/}
\BIBentrySTDinterwordspacing

\bibitem{DockerHub}
\BIBentryALTinterwordspacing
``{Docker Hub}.'' [Online]. Available: \url{https://hub.docker.com/}
\BIBentrySTDinterwordspacing

\bibitem{Nlknguyen/alpine-mpichHub}
\BIBentryALTinterwordspacing
``{nlknguyen/alpine-mpich - Docker Hub}.'' [Online]. Available:
  \url{https://hub.docker.com/r/nlknguyen/alpine-mpich/}
\BIBentrySTDinterwordspacing

\bibitem{MPICHMPI}
\BIBentryALTinterwordspacing
``{MPICH | High-Performance Portable MPI}.'' [Online]. Available:
  \url{http://www.mpich.org/}
\BIBentrySTDinterwordspacing

\bibitem{Heroux2009ImprovingMini-applications}
M.~A. Heroux, D.~W. Doerfler, P.~S. Crozier, J.~M. Willenbring, H.~C. Edwards,
  A.~Williams, M.~Rajan, E.~R. Keiter, H.~K. Thornquist, and R.~W. Numrich,
  ``{Improving Performance via Mini-applications},'' Sandia National
  Laboratories, Tech. Rep. SAND2009-5574, 2009.

\bibitem{CEADevelopment}
\BIBentryALTinterwordspacing
``{CEA - HPC - Research and Development}.'' [Online]. Available:
  \url{http://www-hpc.cea.fr/en/red/red.htm}
\BIBentrySTDinterwordspacing

\bibitem{PerformanceCo-Pilot}
\BIBentryALTinterwordspacing
``{Performance Co-Pilot}.'' [Online]. Available: \url{http://pcp.io/}
\BIBentrySTDinterwordspacing

\bibitem{Sparks2017HPCUse}
\BIBentryALTinterwordspacing
J.~Sparks, ``{HPC Containers in Use},'' pp. 1--8, 2017. [Online]. Available:
  \url{https://cug.org/proceedings/cug2017\_proceedings/includes/files/pap164s2-file1.pdf}
\BIBentrySTDinterwordspacing

\bibitem{Kurtzer2017Singularity:Compute}
\BIBentryALTinterwordspacing
G.~M. Kurtzer, V.~Sochat, and M.~W. Bauer, ``{Singularity: Scientific
  containers for mobility of compute},'' \emph{PLOS ONE}, vol.~12, no.~5, p.
  e0177459, 5 2017. [Online]. Available:
  \url{http://dx.plos.org/10.1371/journal.pone.0177459}
\BIBentrySTDinterwordspacing

\bibitem{Saha2016IntegratingMesos}
\BIBentryALTinterwordspacing
P.~Saha, M.~Govindaraju, S.~Marru, and M.~Pierce, ``{Integrating Apache
  Airavata with Docker, Marathon, and Mesos},'' \emph{Concurrency and
  Computation: Practice and Experience}, vol.~28, no.~7, pp. 1952--1959, 5
  2016. [Online]. Available: \url{http://doi.wiley.com/10.1002/cpe.3708}
\BIBentrySTDinterwordspacing

\bibitem{HomeGateways}
\BIBentryALTinterwordspacing
``{Home - Science Gateways}.'' [Online]. Available:
  \url{https://sciencegateways.org/}
\BIBentrySTDinterwordspacing

\bibitem{Abdelbaky2015DockerCenters}
\BIBentryALTinterwordspacing
M.~Abdelbaky, J.~Diaz-Montes, M.~Parashar, M.~Unuvar, and M.~Steinder,
  ``{Docker Containers across Multiple Clouds and Data Centers},''
  \emph{Proceedings - 2015 IEEE/ACM 8th International Conference on Utility and
  Cloud Computing, UCC 2015}, pp. 368--371, 2015. [Online]. Available:
  \url{https://www.mendeley.com/research-papers/docker-containers-across-multiple-clouds-data-centers/}
\BIBentrySTDinterwordspacing

\bibitem{Nguyen2017DistributedMode}
\BIBentryALTinterwordspacing
N.~Nguyen and D.~Bein, ``{Distributed MPI cluster with Docker Swarm mode},'' in
  \emph{2017 IEEE 7th Annual Computing and Communication Workshop and
  Conference (CCWC)}.\hskip 1em plus 0.5em minus 0.4em\relax IEEE, 1 2017, pp.
  1--7. [Online]. Available: \url{http://ieeexplore.ieee.org/document/7868429/}
\BIBentrySTDinterwordspacing

\bibitem{Ruiz2015PerformanceHPC}
\BIBentryALTinterwordspacing
C.~Ruiz, E.~Jeanvoine, and L.~Nussbaum, ``{Performance Evaluation of Containers
  for HPC}.''\hskip 1em plus 0.5em minus 0.4em\relax Springer, Cham, 8 2015,
  pp. 813--824. [Online]. Available:
  \url{http://link.springer.com/10.1007/978-3-319-27308-2\_65}
\BIBentrySTDinterwordspacing

\bibitem{Bahls2016EvaluatingApplications}
D.~Bahls, ``{Evaluating Shifter for HPC Applications},'' in \emph{Cray User
  Group Conference Proceedings}, 2016.

\bibitem{CraySystems}
\BIBentryALTinterwordspacing
``{Cray XC40 systems}.'' [Online]. Available:
  \url{http://www.cray.com/sites/default/files/resources/cray\_xc40\_specifications.pdf}
\BIBentrySTDinterwordspacing

\bibitem{RktCoreOS}
\BIBentryALTinterwordspacing
``{rkt | rkt Container Engine with CoreOS}.'' [Online]. Available:
  \url{https://coreos.com/rkt}
\BIBentrySTDinterwordspacing

\bibitem{KubernetesOrchestration}
\BIBentryALTinterwordspacing
``{Kubernetes - Production-Grade Container Orchestration}.'' [Online].
  Available: \url{https://kubernetes.io/}
\BIBentrySTDinterwordspacing

\bibitem{Vavilapalli2013ApacheYARN}
\BIBentryALTinterwordspacing
V.~K. Vavilapalli, S.~Seth, B.~Saha, C.~Curino, O.~O'Malley, S.~Radia, B.~Reed,
  E.~Baldeschwieler, A.~C. Murthy, C.~Douglas, S.~Agarwal, M.~Konar, R.~Evans,
  T.~Graves, J.~Lowe, and H.~Shah, ``{Apache Hadoop YARN},'' in
  \emph{Proceedings of the 4th annual Symposium on Cloud Computing - SOCC
  '13}.\hskip 1em plus 0.5em minus 0.4em\relax New York, New York, USA: ACM
  Press, 2013, pp. 1--16. [Online]. Available:
  \url{http://dl.acm.org/citation.cfm?doid=2523616.2523633}
\BIBentrySTDinterwordspacing

\bibitem{Zaharia2016ApacheSpark}
\BIBentryALTinterwordspacing
M.~Zaharia, M.~J. Franklin, A.~Ghodsi, J.~Gonzalez, S.~Shenker, I.~Stoica,
  R.~S. Xin, P.~Wendell, T.~Das, M.~Armbrust, A.~Dave, X.~Meng, J.~Rosen, and
  S.~Venkataraman, ``{Apache Spark},'' \emph{Communications of the ACM},
  vol.~59, no.~11, pp. 56--65, 10 2016. [Online]. Available:
  \url{http://dl.acm.org/citation.cfm?doid=3013530.2934664}
\BIBentrySTDinterwordspacing

\end{thebibliography}

\end{document}